\documentstyle[12pt]{article}

\def\ai{\'{\i}}
\def\itt{\int_{\tau_1} ^{\tau_2}}
\def\qb{\overline Q}
\def\pp{\pi_\phi}

\def\po{\pi_\Omega}
\def\om{\Omega}
\def\33{e^{3\Omega}}
\def\66{e^{6\Omega}}
\def\px{\pi_x}
\def\la{\Lambda}

\def\py{\pi_y}
\def\tst{\textstyle}
\def\e-3{e^{-3\Omega}}

\def\be{\begin{equation}}
\def\ee{\end{equation}}

\def\pb{\overline P}
\begin{document}

\baselineskip.33in

\centerline{\large{\bf Gauge fixation and global phase time for minisuperspaces}}

\bigskip

\centerline{ Claudio Simeone\footnote{Electronic address: simeone@tandar.cnea.gov.ar}}

\medskip

\centerline{\it Departamento de F\ai sica, Comisi\'on Nacional de Energ\ai a At\'omica}

\centerline{\it Av. del Libertador 8250, 1429 Buenos Aires, Argentina}

\centerline{ and}

\centerline{\it Departamento de F\ai sica, Facultad de Ciencias Exactas y Naturales}

\centerline{\it Universidad de Buenos Aires,  Ciudad Universitaria, Pabell\'on I}

\centerline{\it 1428, Buenos Aires, Argentina.}

\vskip1cm

ABSTRACT

\bigskip

Homogeneous and isotropic cosmological models whose Hamilton-Jacobi equation is separable  are deparametrized by turning their action functional into that of an ordinary gauge system. Canonical gauge conditions imposed on the gauge system are used to define a global phase time in terms of the canonical coordinates and momenta of the minisuperspaces.  The procedure clearly shows how the geometry of the constraint surface restricts the choice of time; the consequences that this has on the path integral quantization are discussed.

\vskip1cm

\vskip1cm
\newpage 

\noindent{\bf I. INTRODUCTION}

\bigskip

While in ordinary mechanics the time is an absolute parameter, and this allows for the existence of a unitary quantum theory, in General Relativity the time is an arbitrary label of spacelike hypersurfaces, and physical quantities are invariant under diffeomorfisms. The gravitational field in General Relativity is a parametrized system, its evolution given in terms of a parameter $\tau$ which does not have physical significance.

A possible way to obtain a unitary quantum theory of gravitation is to consider that the time is hidden among the coordinates and momenta of the system, which then must be  deparametrized by identifying the time as a first step before quantization. The identification of time is closely related to  gauge fixation [1]: in the theory of gravitation the dynamical evolution is embodied in the motion of a spacelike hypersurface moving in spacetime along the timelike direction; this motion includes arbitrary local deformations which yield a multiplicity of times. From a different point of view, the same motion can be generated by general gauge transformations. Hence, the gauge fixation is not only a way to select one path from each class of equivalent paths in phase space, but also a reduction procedure identifying a time for the system.

In the present work we  exploit this fact to identify a global phase time [2] for  minisuperspace models whose Hamilton-Jacobi (H-J) equation is solvable. We   define a canonical transformation which turns the cosmological models into ordinary gauge systems by matching their Hamiltonian constraint $H\approx 0$ to one of the new momenta, namely $P_0$, of the gauge system [3,4]. Then we  are able to avoid derivative gauges involving Lagrange multipliers [5,6,7], and to use   gauge conditions given in terms of only the coordinates and momenta ({\it canonical gauges}) to identify a time in terms of the original phase space variables of the cosmological models. The results of ref. 8  are easily reproduced. We  show how the geometry of the constraint surface determines restrictions on the existence of an intrinsic time [9]; we also discuss the consequences that these restrictions have for the path integral quantization of  minisuperspaces,  making more precise the analysis of ref. 3.

\vskip1cm

\noindent{\bf II. PARAMETRIZED SYSTEMS AND ORDINARY GAUGE SYSTEMS}

\bigskip

The action functional of a parametrized system described by the coordinates and momenta $(q^i,p_i)$ has the form
\be
S[q^i,p_i,N]=\itt \left( p_i{dq^i\over d\tau}-NH\right)d\tau\ee
where $N$ is a Lagrange multiplier enforcing the Hamiltonian constraint
\be H(q^i,p_i)\approx 0;\ee
the constraint reflects the reparametrization invariance of the system, i.e. that its evolution is given in terms of the arbitrary parameter $\tau$ which does not have physical meaning.  

The parametrized system described by (1) can be turned into an ordinary gauge system, that is, a system with a true Hamiltonian and a constraint which is linear and homogeneous in the momenta  if the H-J equation is solvable [3]. Consider $W(q^i,\alpha_\mu ,E)$ a complete solution of the $\tau-$independent H-J equation
\be 
H\left( q^i,{\partial W\over\partial q^i}\right) =E\ee
which is obtained by matching the integration constants
 $(\alpha_\mu ,E)$ to  $(\pb_\mu ,\pb_0).$ The solution $W$  generates a canonical transformation
\be 
p_i={\partial W\over\partial q^i},\ \ \ \ \qb^i={\partial W\over\partial \pb_i},\ \ \ \  \overline K=N\pb_0=NH\ee
which identifies the constraint $H$ with the new momentum $\pb_0$. The variables $(\qb^\mu,\pb_\mu)$  are conserved observables because $[\qb^\mu,H]=[\pb_\mu,H]=0$, so that they would not be appropriate to characterize the dynamical evolution. A second transformation generated by the function
\be 
F=P_0\qb^0+f(\qb^\mu ,P_\mu ,\tau) \ee
gives
$$\pb_0={\partial F\over\partial\qb^0}=P_0\ \ \ \ \ \ \  \pb_\mu={\partial f\over \partial\qb^\mu}$$ 
\be Q^0={\partial F\over \partial P_0}=\qb^0\ \ \ \ \ \ \  Q^\mu={\partial f\over\partial P_\mu}\ee 
and a new non vanishing Hamiltonian
\be
K=NP_0+{\partial f\over\partial\tau}=NH+{\partial f\over\partial\tau},\ee
so that $(Q^\mu,P_\mu)$ are non conserved observables.
The two succesive transformations $(q^i,p_i)\ \to\ (\qb^i,\pb_i)\ \to\ (Q^i,P_i)$ lead to the action
\be
{\cal S}=\itt \left( P_i{dQ^i\over d\tau}-NP_0-{\partial f\over\partial\tau}\right) d\tau\ee
which in terms of the original  variables reads [3]
\be
{\cal S}=\itt\left( p_i{dq^i\over d\tau }-NH\right) d\tau +
\left[ \qb^i\pb_i -W+Q^\mu P_\mu-f\right]_{\tau_1}^{\tau_2},\ee
so that $\cal S$ and $S$ differ only in surface terms and then yield the same dynamics. The action (8) contains a linear and homogeneous constraint $P_0\approx 0$ and a non zero Hamiltonian ${{\tst\partial f}\over{\tst\partial\tau}}$ and is then that of an ordinary gauge system.

\vskip1cm

\noindent{\bf III. GAUGE FIXATION AND GLOBAL PHASE TIME}

\bigskip

The constraint $P_0\approx 0$ in equation (8) acts as a generator of gauge transformations yielding an infinite number of physically equivalent paths in the $(Q^i,P_i)$ phase space. To select one path from each class of equivalent paths we must impose a gauge condition $\chi=0$, the choice being restricted by:

\noindent 1) The gauge condition must can be reached from any path by means of gauge transformations leaving the action unchanged.

\noindent 2) Only one point of each orbit (that is, each set of points on the constraint surface  connected by gauge transformations) must be on the manifold defined by $\chi=0$.

To accomplish with 1) the symmetries of the action must be examined; under a gauge transformation generated by a constraint $G$
\be
\delta_\epsilon Q^i=\epsilon (\tau )[Q^i,G], \ \ \ \ \delta_\epsilon P_i=\epsilon (\tau )[P_i,G], \ \ \ \ \delta_\epsilon N={\partial\epsilon (\tau)\over\partial\tau} ,\ee
the variation of the action $\cal S$ is
\be
\delta_\epsilon {\cal S}=\left[ \epsilon (\tau )\left( P_i{\partial G\over\partial P_i}-G\right) \right]_{\tau_1} ^{\tau_2} \ee
and we have $\delta_\epsilon {\cal S}=0$ for $G=P_0$. Therefore the action $\cal S$ is gauge invariant over the whole trajectorie and canonical gauge conditions $\chi(Q^i,P_i,\tau)=0$ are admissible [3]. We should emphasize that if we worked with the original action $S$, as the constraint in $S$ is $H\approx 0$ and $H$ is not linear and homogeneous in the momenta for a parametrized system, then we should fix the gauge by means of a non canonical condition involving a derivative of the multiplier $N$ [5,6]; this is clearly not a good choice if we want to define a global phase time in terms of the phase space variables. 

The condition 2) requires that a gauge transformation moves a point of an orbit off the surface $\chi=0$, so that [10]
$$\delta_\epsilon  \chi=\epsilon (\tau )[\chi, G]\neq 0$$
unless $\epsilon=0$; this holds if
\be [\chi,G]\neq 0.\ee
As $Q^0$ and $P_0$ are conjugate variables,
\be
[Q^0,P_0]=1\ee
so that a gauge condition of the form 
\be
\chi\equiv Q^0-T(\tau)=0\ee
 with $T$ a monotonic function is a good choice. Strictly speeking, equation (12) only ensures that the orbits are not tangent to the surface $\chi=0$; however, as (14) defines a plane $Q^0=constant$ for each $\tau$, if at any $\tau$ any orbit was intersected more than once (then yielding Gribov copies [10]) at another $\tau$ it should be $[\chi, P_0]=0$. Therefore our gauge fixation procedure avoids the Gribov problem.

Given a parametrized system with coordinates and momenta $(q^i,p_i)$ a smooth function $t(q^i,p_i)$  fulfilling
\be 
[t,H]>0\ee
is a global phase time [2]  for the system, and its values along any classical trajectory can parametrize its evolution. As the Poisson bracket is invariant under a canonical transformation,  from (13) and (15) it follows that a globally good gauge choice given in terms of the coordinate $Q^0$ of the gauge system can be used to define a global phase time $t$  for the parametrized system in terms of the coordinates and momenta $(q^i,p_i)$. In other words, a gauge choice for the gauge system defines a particular foliation of spacetime for the parametrized system. We shall see that for certain minisuperspace models a $\tau-$dependent gauge condition of the form $\chi\equiv Q^0-T(\tau)=0$  defines an extrinsic time, that is, a time which is a function not only of the coordinates $q^i$ but also of the momenta $p_i$, while an intrinsic time, i.e. a function of  the coordinates $q^i$ only, can be defined by means of a gauge condition like $\chi\equiv \eta Q^0 P- T(\tau)$ with  $\eta=\pm 1$ if the potential of the model under consideration has a definite sign; in this situation, the constraint surface splits into two disjoint surfaces, and $\eta$ is determined by the sheet on which the system evolves.

\vskip1cm

\noindent{\bf IV. MINISUPERSPACES}

\bigskip

The action of an homogeneous and isotropic Friedmann-Robertson-Walker (FRW) universe is
\be S=\itt\left( \pp\dot\phi+\po\dot\om-NH\right) d\tau\ee
where $\phi$ is the matter field, $\om=\sqrt{4\over 3\pi {\cal G}}\ln a(\tau)$ with $a(\tau)$ the scale factor in the FRW metric,  and $\pp$ and $\po$ are their conjugate momenta; $N$ is a Lagrange multiplier enforcing the Hamiltonian constraint [11] 
\be H=G(\om)(\pp^2-\po^2)+v(\phi,\om ) \approx 0,\ee
where $G(\om)>0$  and $v(\phi,\om)$ is the potential. Our aim is not to study the separability of the H-J equation in general, but to get a clear understanding of the details and also of the restrictions of deparametrizing minisuperspaces by imposing canonical gauge conditions; then we shall limit our analysis to easily solvable models.

\bigskip

{\bf A. A toy model}

\medskip

 Consider the Hamiltonian constraint
\be 
H=-{1\over 4}e^{-3\om}\po^2 +e^{\om}\approx 0
\ee
which corresponds to an open ``universe'' with null cosmological constant. For this model the authors of ref. 8 found a time of the form
 \be t\sim -e^{-4\om/3}\po
\ee
by matching the model to the parametrized system called ``ideal clock'', whose Hamiltonian is ${\tilde H} =p_t -t^2\approx 0.$ They did it by performing a canonical transformation $(t,p_t) \to (\om,\po)$ and multiplying ${\tilde H}$ by a positive definite function of the form $\sim e^{- \om/3}$ to obtain $H$. Then we shall apply our procedure to the constraint $H'=e^{\om/3}H:$
\be
H'=-{1\over 4}e^{-8\om/3}\po^2 + e^{4\om/3}\approx 0.
\ee
The constraint $H'$ is equivalent to  $H$  because they differ  only in a positive definite factor (see below). The  $\tau-$independent H-J equation associated to the Hamiltonian $H'$ is
\be
-\left({\partial W\over\partial\om}\right)^2+4e^{4\om}=4e^{8\om/3}E\ee
and then matching $E=\pb_0$ we have 
\be
W(\om,\pb_0)=\pm\int 2\sqrt{ e^{4\om}-\pb_0 e^{8\om/3}}\- d\om,\ee
with $+$ for $\po>0$ and $-$ for $\po<0$. According to equation (6), on the constraint surface 
\be
Q^0=\qb^0=\left[{\partial W\over\partial\pb_0}\right]_{\pb_0=0}=\mp  e^{2 \om/3}.
\ee
The system described by $Q^0$ and $P_0$ has a constraint which is linear and homogeneous in the momenta.  Its action functional is then invariant under general gauge transformations, so that there is gauge freedom at the endpoints and canonical gauges are admissible. If we choose  $\chi\equiv Q^0-T(\tau)=0$ with $T$ a monotonic function of $\tau$ then we have a global phase time that can be writen in terms of the coordinate $\om$ only, the expression given by the sheet of the constraint surface on which the system evolves: 
\begin{eqnarray} t(\om) &= & -e^{2\om/3}\ \ \ \ \ \ \mbox{if}  \ \ \ \ \ \ \po>0\nonumber\\
t(\om)& = & +e^{2\om/3}\ \ \ \ \ \  \mbox{if}  \ \ \ \ \ \ \po<0.\end{eqnarray}
As on the constraint surface we have 
\be\po=\pm 2e^{2\om},\ee
 then we can write
 \be t(\om,\po)= -{1\over 2}e^{-4\om/3}\po,
\ee
which clearly agrees with (19).

\newpage
\bigskip
{\bf B. True degrees of freedom}
\medskip

Let us go back to the general constraint (17). We shall restrict our analysis to the cases in which the  potential
$  v(\phi,\om)$ has a definite sign. As the cases $v>0$ and $v<0$ are formally analogous, to simplify the notation we shall consider only $v>0$. Define the coordinates
\be 
x=x(\phi+\om),\ \ \ \ \ \ y=y(\phi-\om)\ee
so that ${{\tst \partial x}\over{\tst\partial\phi}}={{\tst\partial  x}\over{\tst \partial\om}}$, ${{\tst \partial  y}\over{\tst \partial\phi}}=-{{\tst \partial y}\over{\tst \partial\om}}$. The momenta $\pi_x,\ \pi_y$ are given by
\be\pi_\phi={\partial x\over\partial\phi}\pi_x+{\partial y\over\partial\phi}\pi_y,\ \ \ \ \ \ \ \ \ \ 
 \pi_\om={\partial x\over\partial\om}\pi_x+{\partial y\over\partial\om}\pi_y,\ee
and then $\pp^2-\po^2=4{{\tst\partial x}\over{\tst\partial\phi}}{{\tst\partial y}\over{\tst \partial\phi}}\pi_x\pi_y=
-4{{\tst\partial x}\over{\tst\partial\om}}{{\tst \partial y}\over{\tst \partial\om}}\pi_x\pi_y.$ If it is possible to choose the coordinates $x$ and $y$ so that $4{{\tst\partial x}\over{\tst\partial\phi}}{{\tst\partial y}\over{\tst \partial\phi}}={{\tst v}\over {\tst G}}$, as ${{\tst v}\over {\tst G}}>0$ then we can multiply the constraint $H$ by $\left( 4G{{\tst\partial x}\over{\tst\partial\phi}}{{\tst\partial y}\over{\tst \partial\phi}}\right)^{-1}$ and obtain a constraint $H'$ which is equivalent to $H$ because it differs only in a positive definite factor:
 \be
H'=\px\py+ 1\approx 0.
\ee

We shall turn the system described by $(x,y,\px,\py)$ into an ordinary gauge system. The $\tau-$independent H-J equation for the constraint (29) is
$${\partial W\over\partial x}{\partial W\over\partial y} + 1=E'$$ and matching the integration constants $\alpha,E'$ to the new momenta $\pb,\pb_0$ it has the solution
\be
W(x,y,\pb_0,\pb)=\pb x+y\left({\pb_0 - 1\over \pb}\right);\ee
then
$$\px={\partial W\over\partial x}=\pb,\ \ \ \ \ \py={\partial W\over\partial y}={\pb_0- 1\over \pb}$$
\be\qb^0={\partial W\over\partial\pb_0}={y\over\pb},\ \ \ \ \ \qb= {\partial W\over\partial\pb}=x+y\left({1 -\pb_0\over\pb^2}\right) .\ee
To go from the set $(\qb^i,\pb_i)$ to $(Q^i,P_i)$ we define
\be
F=\qb^0 P_0 + \qb P+{T(\tau)\over P}\ee
with $T(\tau)$ a monotonic function (see section D for a discussion about this choice). Then we have the canonical variables of the gauge system in terms of those of the minisuperspace:
$$P_0=\px\py+1,\ \ \ \ \ \ P=\px,$$
\be
Q^0={y\over P},\ \ \ \ \ \ Q=x+\left({y(1- P_0)-T(\tau)\over P^2}\right). \ee
There is no problem with $P$ as a denominator because $P=\px$ cannot be zero on the constraint surface.

As $[Q^0,P_0]=1$ we have $[y/\px,H']=1$;  $H'$ differs from $H$ in a positive definite factor, namely $a$, so that $1=[y/\px,H']= [y/\px,aH]=[y/\px,a]H+[y/\px,H]a\approx [y/\px,H]a;$ hence
\be
[y/\px,H]>0\ee
and  a canonical gauge condition of the form $\chi\equiv Q^o-T(\tau)=0$ with $T$ a monotonic function of $\tau$, when imposed on the gauge system described by $Q^i$ and $P_i$ defines a global phase time  $t\equiv y/\px$ for the minisuperspace described by $\phi, \om, \pp, \po .$ From (28) we have $\px=(\pp+\po)\left( 2{{\tst\partial x}\over{\tst\partial\phi}}\right)^{-1}$ and therefore
\be
t(\phi, \om, \pp, \po )=2{y(\phi-\om)\over\pp+\po}{\partial x(\phi+\om)\over\partial\phi}.\ee
The monotonic function of $\tau$ given by (35) depends on the coordinates and also on the momenta of the cosmological model, and is then an extrinsic time.

We can also identify a time in terms of the coordinates only, but, as we shall see, the definition  depends on the sheet of the constraint surface on which  the system evolves. The gauge choice
\be 
\chi\equiv\eta Q^0P-T(\tau)=0
\ee
with $\eta=\pm 1$ gives
\be
[\chi, P_0]=\eta P
\ee
and as $\eta Q^0P=\eta y$ and $P=\px$ we have
\be
[\eta y, H']=\eta \px.
\ee
As before, as $H'$ and $H$ differ in a positive definite factor, if we can define $\eta$ so that $[\eta y, H']>0$ then $[\eta y, H]>0$ and $\eta y$ is a global phase time. We can  chose ${{\tst\partial x}\over{\tst\partial\phi}}$ as a positive definite function (and appropriately adjust the sign of ${{\tst\partial y}\over{\tst\partial\phi}}$) to yield $sign (\px)=sign (\pp+\po)$. From the constraint equation we have
\be
\po=\pm\sqrt{{v(\phi,\om)\over G(\om)}+\pp^2}\ee
and because ${v/G}$ is positive definite, $\po\neq 0$ and the evolution of the system is restricted to one of the two disjoint surfaces (39), each one topologically equivalent to half a plane. Moreover, from (39) we have $|\po| > |\pp|$, yielding $sign(\px) >0$ for $\po > 0$ and $sign(\px) <0$ for $\po < 0$. 
Hence we can have a good definition of time on each sheet of the constraint surface by appropriately choosing $\eta$, the choice dictated by the sign of the momentum $\po :$ 
\begin{eqnarray}t(\phi,\om) & = & +y(\phi-\om)\ \ \ \ \ \ \mbox{if}  \ \ \ \ \ \ \po>0\nonumber\\
t(\phi,\om)& = & -y(\phi-\om)\ \ \ \ \ \  \mbox{if}  \ \ \ \ \ \ \po<0.\end{eqnarray}

Therefore,   even though we can not write a single expression  which holds for both sheets of the constraint surface, if $v$ has a definite sign, once we have on which sheet the system evolves we can identify a time in terms of  the coordinates  (intrinsic time). If, instead, we want an expression which holds automatically, that is, which does not depend on the sign of $\po$, we must choose a time like that given in (35).

\newpage
\bigskip

\noindent{\it Examples}

\medskip
 
\noindent 1) Consider a flat model with massless scalar field $\phi$ and a cosmological ``constant'' which decays with $\phi$ as $\Lambda=\Lambda_0 e^{-6\phi}$:
\be H={1\over 4}e^{-3\om}(\pp^2-\po^2)+ \Lambda_0 e^{-6\phi} e^{3\om}\approx 0.\ee
This constraint is equivalent to
$$H'=\pp^2-\po^2+4\Lambda_0\ e^{-6(\phi-\om)}\approx 0,$$
making the choice of variables $x=\phi+\om$, $y=-{{\tst \Lambda_0}\over {\tst6}} e^{-6(\phi-\om)}$ obvious; by turning the system into an ordinary gauge system with coordinates and momenta $(Q^0,Q,P_0,P)$ and fixing the gauge with the canonical condition $\chi\equiv Q^0-T(\tau)=0$ with $T$ a monotonic function we obtain the time
\be
t(\phi,\om,\pp,\po)=-{1\over 3}{\Lambda_0 e^{-6(\phi-\om)}\over \pp+\po},\ee
which on the constraint surface is equivalent to
$$t(\pp,\po)={1\over 2}(\pp-\po).$$
The system also has an intrinsic time, which according to (40) can be writen as
$$t=\mp{{\tst \Lambda_0}\over {\tst6}} e^{-6(\phi-\om)},$$
with $-$ if the system is on the sheet $\po>0$ and $+$ if it is on the sheet $\po<0$.

\medskip 

\noindent 2) A closed $(k=1)$ model with cosmological constant $\Lambda>0$ and massless scalar field $\phi$, whose Hamiltonian constraint is
\be
H={1\over 4}e^{-3\om}(\pp^2-\po^2)-e^\om+ \Lambda e^{3\om}\approx 0\ee
is not separable in terms of the variables $x(\phi+\om),\ y(\phi-\om)$; moreover, its potential has not a definite sign. However, it is easy to show that the time obtained for the case $k=0$ (flat model) is also a global phase time for the case 
$k=1$. Then consider the constraint
\be
H_0={1\over 4}e^{-3\om}(\pp^2-\po^2)+\Lambda e^{3\om}\approx 0\ee
which is equivalent to
$$H_0'=\pp^2-\po^2+4\Lambda e^{6\om}\approx 0.$$
By choosing $y=-(1/3)e^{3(\om-\phi)},\ x=(1/3)e^{3(\om+\phi)}$ the  same procedure used in the preceding example gives the extrinsic time
\be
t=-2/3{\la e^{6\om}\over \pp+\po}\approx {1\over 6}(\pp-\po).\ee
Note that if we want to verify that this function is a global phase time also for the case $k=1$ we should not write it as ${1\over 6}(\pp-\po)$ because the last equality  holds only on the surface  $H_0\approx 0$. If we calculate the Poisson bracket of $t=-2/3{ {\tst \la e^{6\om}}\over{\tst \pp+\po}}$ with the constraint $H'=4e^{3\om} H$ we  obtain $[t,H']=[t,H_0']+[t,-4e^{4\om}]$, which, as it is easy to check, is the sum of two positive terms. As the constraints $H$ and $H'$ are equivalent, then we have
$$[t,H]>0$$
and $t$ is a global phase time also for the model given by (43).

\bigskip
{\bf C. Geometry of the constraint surface}
\medskip

Our deparametrization procedure gives a simple way to 
examine how the geometrical properties of the constraint surface imposes restrictions on the definition of a global phase time. Consider the Hamiltonian constraint of the most general case of a FRW empty cosmological model:
\be 
H=-{1\over 4}e^{-3\om}\po^2 -ke^{\om}+\Lambda e^{3\om}\approx 0
\ee
with $k=\pm 1$  and $\la >0$.  For this model the authors of ref. 8 found an extrinsic time
\be t\sim -e^{-2\om}\po
\ee
after performing a canonical transformation on the   ideal clock and multiplying the constraint by a positive definite function of the form $\sim e^{ \om}$.  Then we shall apply our procedure to the constraint $H'=e^{-\om}H:$
\be
H'=-{1\over 4}e^{-4\om}\po^2 -k+\la e^{2\om}\approx 0.
\ee
The constraints $H$ and $H'$ are equivalent because they differ  only in a positive definite factor. The  $\tau-$independent H-J equation for the Hamiltonian $H'$ is
\be
-\left({\partial W\over\partial\om}\right)^2-4ke^{4\om}+4\la e^{6\om}=4e^{4\om}E\ee
and matching $E=\pb_0$ we obtain the solution
\be
W(\om,\pb_0)=\pm\int 2e^{2\om}\sqrt{\la e^{2\om}-k-\pb_0}\-d\om,\ee
with $+$ for $\po>0$ and $-$ for $\po<0$. According to equation (6), on the constraint surface we have
\be
Q^0=\qb^0=\left[{\partial W\over\partial\pb_0}\right]_{\pb_0=0}=\mp \la^{-1}\sqrt{\la e^{2 \om}-k}.
\ee
If we fix the gauge by means of the canonical condition $\chi\equiv Q^0-T(\tau)=0$ with $T$ a monotonic function of $\tau$ then we have that 
\be t= \theta(-\po)\ \la^{-1}\sqrt{\la e^{2 \om}-k}-\theta(\po)\ \la^{-1}\sqrt{\la e^{2 \om}-k}\ee
is a global phase time for the system. 
As on the constraint surface we have 
\be\po=\pm 2e^{2\om}\sqrt{\la e^{2\om}-k},\ee
(so that in the case $k=1$ the natural size of the configuration space is given by $\om\geq -\ln(\sqrt{\la})$ [2])
 then we can write
 \be t(\om,\po)= -{1\over 2}\la^{-1}e^{-2\om}\po,
\ee
which is in agreement with (47). Now an important difference between the cases $k=-1$ and $k=1$ arises: for $k=-1$ the potential has a definite sign, and the constraint surface splits into  two disjoint sheets given by (53). In this case the evolution can be parametrized by a function of the coordinate $\om$ only, the choice given by  the sheet on which the system remains, and we then say that it has an intrinsic time: if the system is on the sheet $\po>0$ the time is $t= -\la^{-1}\sqrt{\la e^{2 \om}-k}$, and if it is on the sheet $\po<0$ we have $t=\la^{-1}\sqrt{\la e^{2 \om}-k}$.
For $k=1$, instead, the potential can be zero and the topology of the constraint surface is no more analogous to that of two disjoint planes. Although for $\om= -\ln(\sqrt{\la})$ we have $v(\om)=0$ and $\po=0,$ it is easy to verify that ${\dot \po}\neq 0$ at this point.   Hence, in this case  the coordinate $\om$ does not suffice to parametrize the evolution, because the system can go from $(\om,\po)$ to $(\om,-\po)$; therefore we must necessarily define a global phase time as a function of the coordinate and the momentum (extrinsic time): $t=t(\om,\po)$. This, of course, generalizes to the case of models with true degrees of freedom.

\medskip

A  remark should be made, and it is that we have multiplied $H$ by different positive functions to make calculations simpler, or to obtain times that we could compare with previous results;  different rescalings of the Hamiltonian constraint would lead to different times, but --at least at the classical level-- they would be equivalent.

\bigskip
{\bf D. Path integral quantization}
\medskip

Suppose that we want to quantize a cosmological model described by $(q^i,p_i)$ by means of a path integral in terms of the variables $(Q^i,P_i)$ given by (33).
 As we showed in a previous paper [4], this has practical advantages, for example, when trying to avoid the Gribov problem. If we pretend the quantum amplitude $<Q_2^i|Q_1^i>$ to be equivalent to $<q_2^i|q_1^i>$  we should verify that the paths in the integral are weighted by the action ${\cal S}$ in the same way that they are weighted by $S$, and that the quantum states $|Q^i>$ are equivalent to $|q^i>.$ As the path integral in the variables $(Q^i,P_i)$ is gauge invariant, this requirement is fulfilled if it is possible to impose a -globally good- gauge condition $\tilde \chi=0$ such that $\tau=\tau(q^i)$ is defined, and such that the boundary terms in (9) vanish.  This is the reason why we chose the generating function for $\overline X^i \to X^i$ as in (32): with this choice the boundary terms in (9) vanish if we fix the gauge by means of $\chi\equiv\eta PQ^0-T(\tau)=0$; when writen in terms of the original variables,  this gauge condition involves only the coordinates $q^i$, and is associated to the identification of an intrinsic time.

An intrinsic time, however, can be defined  only if the constraint surface splits into two disjoint sheets, that is, if the potential has a definite sign. In the most general case the definition of a global phase time must necessarily involve also the momenta, and then we cannot fix the gauge in the path integral in such a way that $\tau=\tau(q^i)$ (see the last example, where $t= -1/2\la^{-1}e^{-2\om}\po\approx T(\tau),$ so that $\tau=\tau(\om,\po)$). Hence, if we want to quantize the system by imposing canonical gauges in the path integral, in the most general case of a potential with a non definite sign  we must admit the possibility of identifying the quantum states in the original phase space not by $q^i$ but  by a complete set of functions of the coordinates and momenta $q^i$ and $p_i$.

\vskip1cm

\noindent{\bf V. CONCLUSIONS}
 
\bigskip

Although gauge fixation and  the identification of a global phase time are closely related,  as the action of parametrized systems --like the gravitational field-- is not gauge invariant at the boundaries, we could not, in principle,  use this fact to obtain a direct procedure to deparametrize minisuperspaces: while ordinary gauge systems admit gauge conditions of the type $\chi(q^i,p_i,\tau)=0$,  only derivative gauges would be admissible for parametrized systems. Then we would not be able to identify a time for a cosmological model as a funtion of its canonical variables by imposing on the systen a gauge condition which is compatible with the symmetries of the action. 
 
However, if the H-J  equation is separable, a parametrized system described by $(q^i,p_i)$ can be turned into an ordinary gauge system described by $(Q^i,P_i)$ by matching $H$ with $P_0$, and canonical gauges are therefore admissible. Then we are able to identify a global phase time for cosmological models
in terms of their coordinates and momenta  by imposing $\tau-$dependent canonical gauge conditions
on the ordinary gauge system.   We have illustrated our procedure with simple models whose H-J equation is easily solvable. We have been able to show that sometimes a global phase time for a quite trivial model is also a good time for a more physical system (example 2); however, we believe that this is not the best way to proceed, because in a general case it would only work if we impose restrictions on the parameters of the model (as it happens when we consider a massive scalar field, when a relation between $\la$ and $m$ should exist). Of course, a complete solution of the H-J equation is, in general, difficult to obtain; an example of a more   interesting model to be studied could be  the Bianchi type-IX universe, which is the anisotropic generalization of the closed FRW model, and whose  H-J equation is solvable [12].

Our procedure clearly shows the restrictions arising from the geometry of the constraint surface: a global phase time in terms of the coordinates $q^i$ can be defined only if the potential of the model has a definite sign; in this case, the choice is determined by the sheet of the constraint surface on which the system evolves. In the most general case,  a global phase time must be a function of the coordinates and the momenta; at the quantum level,  our method completes the analysis of ref. 3, clearly showing  the relation existing between the geometrical properties of the constraint surface  and the possibility of identifying the quantum states in the path integral by means of only the original coordinates.

\vskip1cm

REFERENCES

\bigskip

\noindent 1. A. O. Barvinsky, Phys. Rep. {\bf 230}, 237 (1993).

\noindent 2. P. H\'aj\ai cek, Phys. Rev. D {\bf 34}, 1040 (1986).

\noindent 3. R. Ferraro and  C. Simeone,  J. Math. Phys. {\bf 38}, 599 (1997).

\noindent 4. C. Simeone, J. Math. Phys. {\bf 39}, 3131 (1998).

\noindent 5. C. Teitelboim, Phys. Rev. D {\bf 25}, 3159 (1982).

\noindent 6. J. J. Halliwell, Phys. Rev. D {\bf 38}, 2468 (1988).

\noindent 7. M.   Henneaux, C.  Teitelboim and J.  D.  Vergara, Nucl. Phys. B 
{\bf 387}, 391 (1992).

\noindent 8. S.  C.   Beluardi and R.  Ferraro, Phys. Rev. D {\bf 52}, 1963 (1995).

\noindent 9. K. V. Kucha\v r, in {\it Proceedings of the 4th Canadian Conference on General Relativity and Relativistic Astrophysics}, edited by  G. Kunstatter, D. Vincent and J. Williams, World Scientific, Singapore  (1992).

\noindent 10. M.  Henneaux and C.  Teitelboim, {\it Quantization of Gauge Systems,}
Princeton University Press, New Jersey (1992).

\noindent 11.  J.    J.   Halliwell, in {\it Introductory Lectures  on  Quantum
Cosmology,}  Proceedings  of  the Jerusalem Winter School on Quantum Cosmology
and Baby Universes, edited by  T. Piran, World Scientific, Singapore (1990).

\noindent 12. V. Moncrief and M. P. Ryan Jr., Phys. Rev. D {\bf 44}, 2375 (1991).

\end{document}